\documentclass[runningheads]{llncs}
\usepackage[T1]{fontenc}
\usepackage{graphicx}
\usepackage{booktabs}
\usepackage[misc]{ifsym}

\usepackage{mwe}
\usepackage{url}
\usepackage{multirow}
\usepackage{caption}
\usepackage{diagbox}
\usepackage{amsmath}
\usepackage{amssymb}
\usepackage{algorithm}
\usepackage{algpseudocode}
\usepackage{booktabs}
\usepackage{subcaption}
\usepackage{tcolorbox}
\newtheorem{assumption}{Assumption}

\begin{document}

\title{Tail-Aware Adaptive-k: Query-Adaptive Context Selection for Retrieval-Augmented Generation}
\tocauthor{Ziyu Song, Jiaming Fang, Kuangyu Li, Tuo Xia, Chuanpeng Wang}
\toctitle{Tail-Aware Adaptive-k: Query-Adaptive Context Selection for Retrieval-Augmented Generation} 

\titlerunning{Tail-Aware Adaptive-k}

\author{Ziyu Song\inst{1}$^{\star}$ \and
    Jiaming Fang\inst{1,2}$^{\star}$ \and
    Kuangyu Li\inst{1} \and
    Tuo Xia\inst{1}\Letter \and
    Chuanpeng Wang\inst{1}}      
\institute{AI Lab, 37 Interactive Entertainment, Guangzhou, 510300, China \\
\{songziyu, likuangyu, xiatuo, wangchuanpeng\}@37.com \and
School of Computer Science, Wuhan University, Wuhan, 430072, China \\
fang\_jiaming@whu.edu.cn}   

\maketitle              
\footnotetext{$^{\star}$ These authors contributed equally to this work.}

\begin{abstract}
Adaptive context selection is critical for retrieval-augmented generation (RAG) 
systems, as fixed Top-$K$ retrieval fails under query-dependent and heavy-tailed 
similarity distributions. While Extreme Value Theory (EVT) offers a principled 
framework for adaptive truncation, existing approaches apply EVT globally across the entire ranked list, incurring prohibitive computational costs and statistical instability.We propose \textbf{Tail-Aware Adaptive-$k$ (TAA-$k$)}\footnote{\url{https://anonymous.4open.science/r/pkdd2026taak}}, a training-free framework that operationalizes EVT through a \emph{localized validation} strategy. The key insight is that ranked similarity curves exhibit a characteristic steep--flat--steep 
pattern reflecting a transition from relevance-dominated to noise-dominated regimes. TAA-$k$ exploits this geometric structure via knee detection to identify a compact candidate region, then applies EVT-based goodness-of-fit testing within this window to validate the onset of tail behavior. This coarse-to-fine design reduces computational complexity from $O(N^2M)$ 
to $O(\sqrt{N \log N} \cdot M)$ while maintaining statistical rigor. Under mild monotone likelihood ratio assumptions, TAA-$k$ yields a stable, query-adaptive cutoff corresponding to the earliest noise-dominated position. Experiments on WebQuestions, 2WikiMultiHopQA, and MuSiQue demonstrate that TAA-$k$ achieves near-oracle retrieval quality (F1 within 2--3\% of oracle) with orders-of-magnitude efficiency gains over global EVT methods, while maintaining robustness across embedding models and compression dimensions.
\end{abstract}

\section{Introduction}
Retrieval-Augmented Generation (RAG) has become a central paradigm for enhancing large language models with external knowledge, enabling more factual, up-to-date, and verifiable generation \cite{lewis2020retrieval,gao2023retrieval}. 
In a typical RAG pipeline, a retriever ranks candidate documents or passages according to their similarity to a query, and a subset of the ranked results is selected as contextual input for generation. 
The effectiveness of this context selection step is critical: retaining too many low-relevance items introduces noise and degrades generation quality, while overly aggressive truncation risks discarding essential evidence, especially for multi-hop or compositional queries \cite{jiang2023active}.

\begin{figure}[ht]
\centering
\includegraphics[height=4cm]{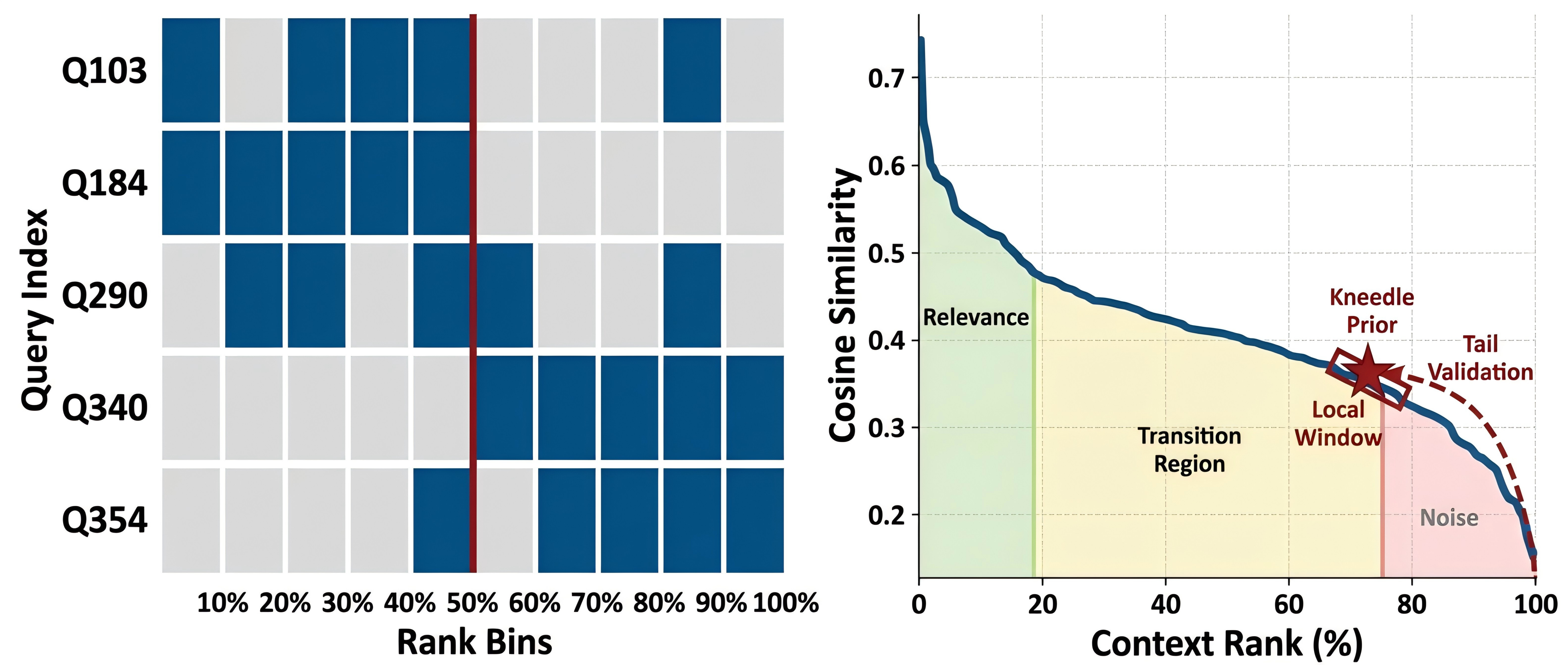}
\caption{
Overview of the proposed Tail-Aware Adaptive-$k$ (TAA-$k$) framework. \textbf{Left.} Relevant contexts (blue) are unevenly distributed across rank bins for different queries, while a fixed cutoff (red) either includes low-relevance contexts or excludes informative ones, demonstrating the instability of fixed Top-$K$ selection as corpus size and heterogeneity increase. \textbf{Right.} Ranked similarity curves typically exhibit a characteristic \emph{steep--flat--steep} pattern, corresponding to a relevance-dominated head, a transition region, and a noise-dominated tail. 
Leveraging this structure, TAA-$k$ identifies the truncation boundary via a coarse-to-fine strategy: knee detection first locates a candidate transition region, followed by localized EVT-based tail stability validation.
}
\label{fig:overview}
\end{figure}

Despite its importance, most RAG systems rely on a fixed Top-$K$ retrieval strategy \cite{lewis2020retrieval,karpukhin2020dense}, implicitly assuming that similarity score distributions are stable across queries and corpora. 
As shown in Fig.~\ref{fig:overview}(left), empirical evidence suggests otherwise: retrieval scores exhibit strong query-dependent variability, particularly in large and heterogeneous corpora.
As a result, a single global cutoff is inherently brittle, frequently including irrelevant contexts for some queries while excluding critical information for others.

Fig.~\ref{fig:overview}(right) illustrates a typical ranked similarity curve 
from dense retrieval exhibiting a characteristic \emph{steep--flat--steep} pattern. 
This reflects a transition between relevance regimes: a \textit{relevance-dominated head}, 
a \textit{transition region}, and a \textit{noise-dominated tail}. 
The optimal truncation boundary typically lies near the tail onset.

Recent work has explored dynamic or heuristic truncation strategies to mitigate this issue, including gap-based rules \cite{taguchi-etal-2025-efficient}, confidence-based filtering, and model-driven selection mechanisms \cite{bahri2023surprise,li2024retrieval}. 
While these approaches improve over fixed Top-$K$, they either rely on manually tuned heuristics, incur substantial computational overhead, or depend on additional learned models, limiting their robustness and scalability.

From a statistical perspective, the tail behavior of ranked similarity scores provides a natural signal for adaptive truncation. 
Extreme Value Theory (EVT) \cite{pickands1975statistical} offers principled tools for modeling distributional extremes and has recently been applied to result list truncation \cite{bahri2023surprise}. 
However, existing EVT-based approaches typically perform global tail fitting across the entire ranked list, leading to prohibitive $O(N^2)$ complexity and instability under finite samples. 
Moreover, in practical retrieval scenarios, the utility of EVT lies less in asymptotic tail modeling than in diagnosing the onset of a noise-dominated regime.

In this work, we propose \textbf{Tail-Aware Adaptive-$k$ (TAA-$k$)}, a training-free and query-adaptive context selection framework that combines geometric structure in ranked similarity curves with localized statistical tail validation. 
As illustrated in Fig.~\ref{fig:overview}(right), our method follows a coarse-to-fine strategy. 
First, we exploit the geometric structure of the ranked similarity curve by applying knee detection to identify a compact candidate region where the relevance-to-noise transition is likely to occur. 
Second, we perform localized EVT-inspired goodness-of-fit testing within this region to validate the stability of the tail distribution. 
The truncation boundary is determined as the earliest index at which tail stability is observed.

Importantly, our approach does not assume that similarity scores strictly follow an EVT limit distribution. 
Instead, Generalized Pareto models are used as a diagnostic tool: once the ranked list enters a noise-dominated regime, the fitted tail statistics tend to stabilize under further truncation. 
This geometry-guided, localized design dramatically reduces computational cost while preserving the ability to identify query-specific truncation boundaries adaptively.

We evaluate TAA-$k$ on three standard RAG benchmarks—WebQuestions \cite{berant2013semantic}, 2WikiMultiHopQA \cite{ho2020constructing}, and MuSiQue \cite{trivedi2022musique}—and demonstrate that it achieves retrieval quality close to an oracle truncation strategy, while being orders of magnitude more efficient than global EVT-based methods. 
Extensive experiments across retrieval models and embedding dimensions further show that our method is robust and model-agnostic.

Our contributions are summarized as follows:
\begin{itemize}
    \item We formulate adaptive context selection in RAG as a tail stability detection problem on ranked similarity scores, enabling query-specific truncation without additional training.
    \item We introduce a geometry-guided, localized tail validation framework that integrates knee detection with EVT-inspired goodness-of-fit testing, substantially reducing computational overhead.
    \item We empirically demonstrate that the proposed method achieves near-oracle retrieval quality with strong robustness across datasets, retrievers, and embedding dimensions.
\end{itemize}

\section{Related Work}

\subsection{Retrieval-Augmented Generation}

Retrieval-Augmented Generation (RAG) has emerged as a prominent paradigm for augmenting large language models with external knowledge sources. Early RAG systems combine neural retrievers with sequence-to-sequence generators \cite{lewis2020retrieval}, with subsequent improvements through dense representation learning \cite{karpukhin2020dense} and hybrid retrieval pipelines \cite{sawarkar2024blended}. However, an increasingly recognized bottleneck concerns how many retrieved contexts should be supplied to the generator. Empirical studies show that long contexts degrade model performance due to attention dilution, positional bias, and evidence interference \cite{liu2024lost}, making adaptive context selection a critical component of practical RAG pipelines.

\subsection{Adaptive Context Selection in RAG}

Most RAG implementations adopt fixed Top-$K$ retrieval, assuming stable similarity score distributions across queries and corpora. However, empirical analyses demonstrate strong query-dependent variability in ranking scores, particularly in large heterogeneous corpora, making global cutoffs unreliable.

Existing adaptive strategies fall into two categories. Heuristic approaches like Adaptive-$k$ \cite{taguchi-etal-2025-efficient} estimate truncation points using score gaps or local ranking statistics, offering computational efficiency but suffering from sensitivity to score scaling and poor performance under relevance overlap. Model-based approaches such as Self-Route \cite{li2024retrieval} employ learned confidence estimation or LLM-driven selection, improving precision but incurring substantial computational overhead and requiring additional supervision.

In contrast, we formulate context truncation as a statistical regime detection problem over ranked similarity scores, enabling query-adaptive truncation without additional training while explicitly modeling the transition from relevance-dominated to noise-dominated regimes.

\subsection{Extreme Value Theory and Tail Diagnostics for Ranking}

Extreme Value Theory (EVT) \cite{pickands1975statistical} provides a principled framework for modeling distributional extremes and threshold selection \cite{coles2001introduction,embrechts2013modelling}. In information retrieval, EVT has been applied to characterize ranking distributions and guide result truncation. Most closely related is Surprise \cite{bahri2023surprise}, which applies global EVT-based tail modeling via Generalized Pareto Distributions (GPDs) across candidate truncation points. While theoretically grounded, this global approach incurs quadratic computational cost and suffers statistical instability when tails contain relevant items.

Our approach differs fundamentally: we treat GPD fitting as a diagnostic tool for detecting noise-dominated regimes rather than assuming global modeling. Critically, we integrate EVT diagnostics with geometric structure in ranked similarity curves. By exploiting the characteristic steep--flat--steep pattern, we first localize a compact candidate region via knee detection, then perform EVT-based goodness-of-fit testing only within this region. This geometry-guided localization transforms global EVT modeling into a localized tail diagnostic procedure, substantially reducing the search space while preserving statistical interpretability.

\section{Method}

\subsection{Problem Setup}

Let $\mathcal{D}=\{d_1,\dots,d_N\}$ be a set of retrieved documents
associated with similarity scores
$S=\{s_1 \ge s_2 \ge \dots \ge s_N\}$,
sorted in descending order.

Our goal is to determine an adaptive truncation index $k^*$
such that:
\begin{itemize}
    \item the prefix $\{s_1,\dots,s_{k^*}\}$ is dominated by relevant items;
    \item the suffix $\{s_{k^*+1},\dots,s_N\}$ constitutes
    a statistically stable noise tail.
\end{itemize}

Formally, we seek the smallest index $k^*$
such that the conditional distribution of
$\{s_i : i > k^*\}$
is asymptotically invariant under further truncation,
indicating entry into a noise-dominated regime.

This problem can be formulated as detecting the onset of a noise-dominated tail regime in a ranked similarity sequence.


\subsection{Empirical Structural Observation}

Across retrieval models and datasets,
ranked similarity curves consistently exhibit
a \emph{steep--flat--steep} pattern.

We normalize the ranked similarity curve by
\begin{equation}
x_i = \frac{i}{N},
\qquad
y_i = \frac{s_i - s_N}{s_1 - s_N},
\end{equation}
yielding a monotone decreasing curve $(x_i,y_i)\in[0,1]^2$.

Empirically, the curve often shows a concave--convex transition,
which is consistent with order statistics of a two-component mixture
under monotone likelihood ratio ordering:
high-similarity ranks are relevance-dominated,
while the lower tail concentrates noise.
This geometric transition motivates
a statistical characterization of the tail regime.


\subsection{Mixture Model and Identifiable Tail Regime}

We model similarity scores as samples from a mixture \cite{robertson2009probabilistic}:
\begin{equation}
p(s) = \pi_r p_r(s) + (1-\pi_r) p_t(s),
\end{equation}
where $p_r$ denotes the relevance distribution
and $p_t$ denotes the noise (tail) distribution.

\begin{assumption}[Sufficient Condition for Identifiable Tail]
\label{ass:tail_sep}
Assume that:
\begin{enumerate}
    \item $p_r(s)$ and $p_t(s)$ are continuous densities on $[0,1]$;
    \item the likelihood ratio
    $\Lambda(s) := p_r(s)/p_t(s)$
    is strictly increasing in $s$
    (monotone likelihood ratio property).
\end{enumerate}
Then there exists at most one score $s_c$
such that $p_r(s_c)=p_t(s_c)$.
For all $s < s_c$, the mixture is tail-dominated in the sense that
$p_t(s) \ge p_r(s)$.
\end{assumption}

Our Assumption~\ref{ass:tail_sep} is consistent with classical probabilistic relevance models in information retrieval \cite{robertson2009probabilistic}, and
 Assumption~\ref{ass:tail_sep} serves as a sufficient rather than a necessary condition.
Under this assumption,
there exists a unique population-level transition
between relevance-dominated and noise-dominated regimes.
Let $k_c$ denote the index corresponding to $s_c$
in the ranked sequence.

Our method does not estimate mixture parameters explicitly. Instead, it identifies the earliest index
beyond which the ranked samples
behave as if generated solely from the noise distribution.
When the assumption is mildly violated,
the proposed approach remains a data-driven heuristic.


\subsection{Lower-Tail Modeling via Extreme Value Theory}

Similarity scores are bounded in practice,
and noise corresponds to low similarity values.

For a candidate truncation index $k$,
define a lower threshold $l_k := s_k$
and reflected tail exceedances:
\begin{equation}
z_i = s_i - l_k, \quad i > k.
\end{equation}
Let $Z_k$ denote the distribution of $z_i$
conditional on $s_i < l_k$.

\paragraph{Extreme Value Approximation.} We assume that the noise distribution $p_t(s)$
has a finite lower endpoint
and belongs to the Weibull maximum domain of attraction,
which is standard for bounded similarity measures.
Under this condition,
for any sequence of thresholds $l_k$
such that $|T_k|\to\infty$ and $l_k$ approaches the lower endpoint,
the conditional distribution of $Z_k$
converges to a Generalized Pareto Distribution (GPD)
with shape parameter $\xi<0$.

Importantly, we do not rely on asymptotic exactness.
Our criterion depends on \emph{relative tail stability}:
once tail samples are dominated by $p_t$,
the fitted GPD parameters and goodness-of-fit statistics
vary slowly with further truncation.

When $k < k_c$,
the tail contains a non-vanishing fraction of samples from $p_r$,
which induces deviations from the GPD family
and inflates goodness-of-fit statistics.
Such contamination effects are well documented
in EVT-based threshold selection.


\subsection{Tail Stability Criterion}

For each candidate index $k$,
define the tail set:
\begin{equation}
T_k = \{ s_i : i > k \}.
\end{equation}

We fit a GPD to the reflected exceedances
using maximum likelihood estimation,
and evaluate goodness of fit via
the Cramér--von Mises (CVM) \cite{cramer1928composition,anderson1952asymptotic} statistic:
\begin{equation}
\mathrm{CVM}(k)
=
\int
\left[
\hat F_k(x) -
G_{\hat\xi_k,\hat\sigma_k}(x)
\right]^2
\, dG_{\hat\xi_k,\hat\sigma_k}(x),
\end{equation}
where $\hat F_k$ is the empirical CDF of $Z_k$. 

\begin{figure}[ht]
  \centering
  \begin{minipage}{0.48\linewidth}
    \centering
    \includegraphics[width=\linewidth]{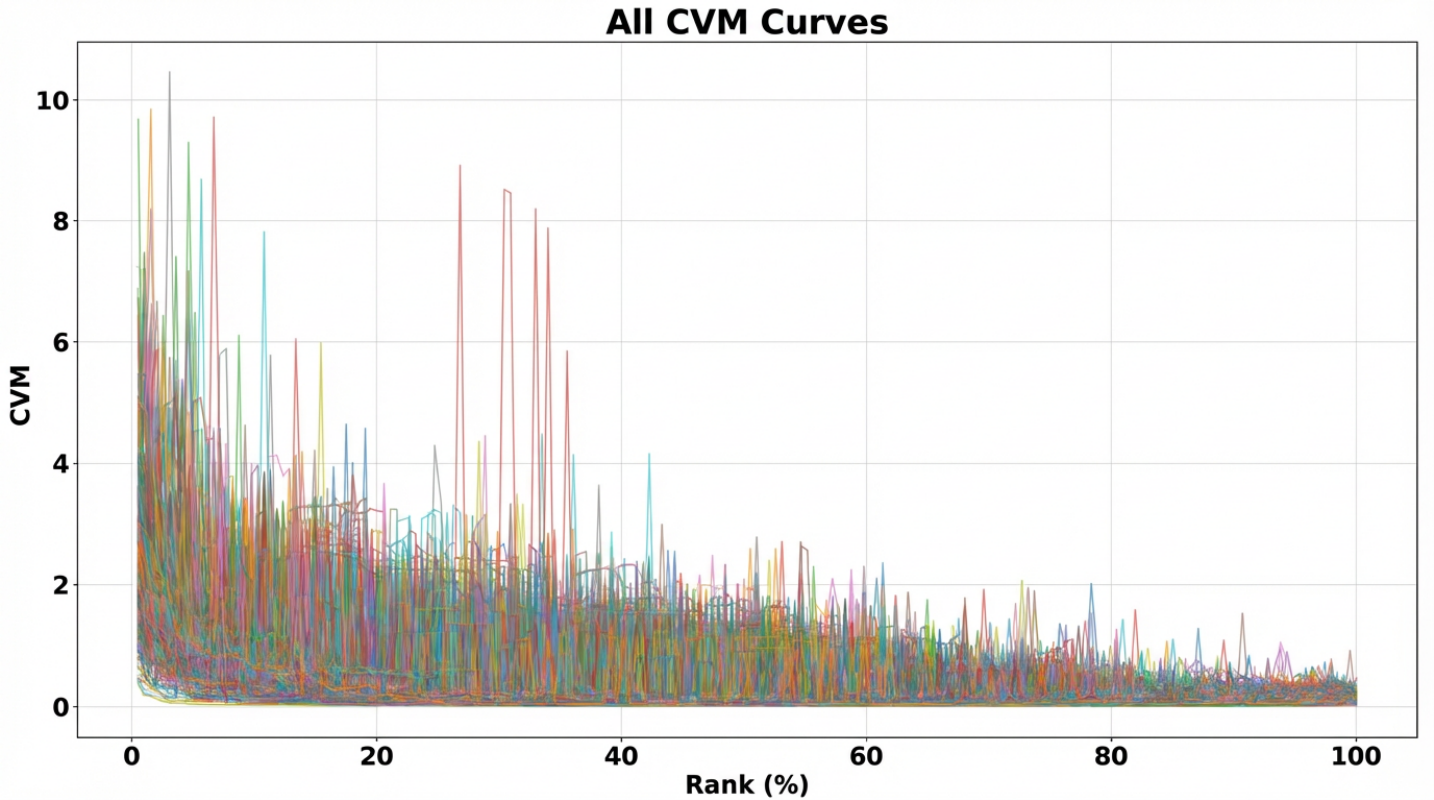}
    \small (a)
  \end{minipage}\hfill
  \begin{minipage}{0.46\linewidth}
    \centering
    \includegraphics[width=\linewidth]{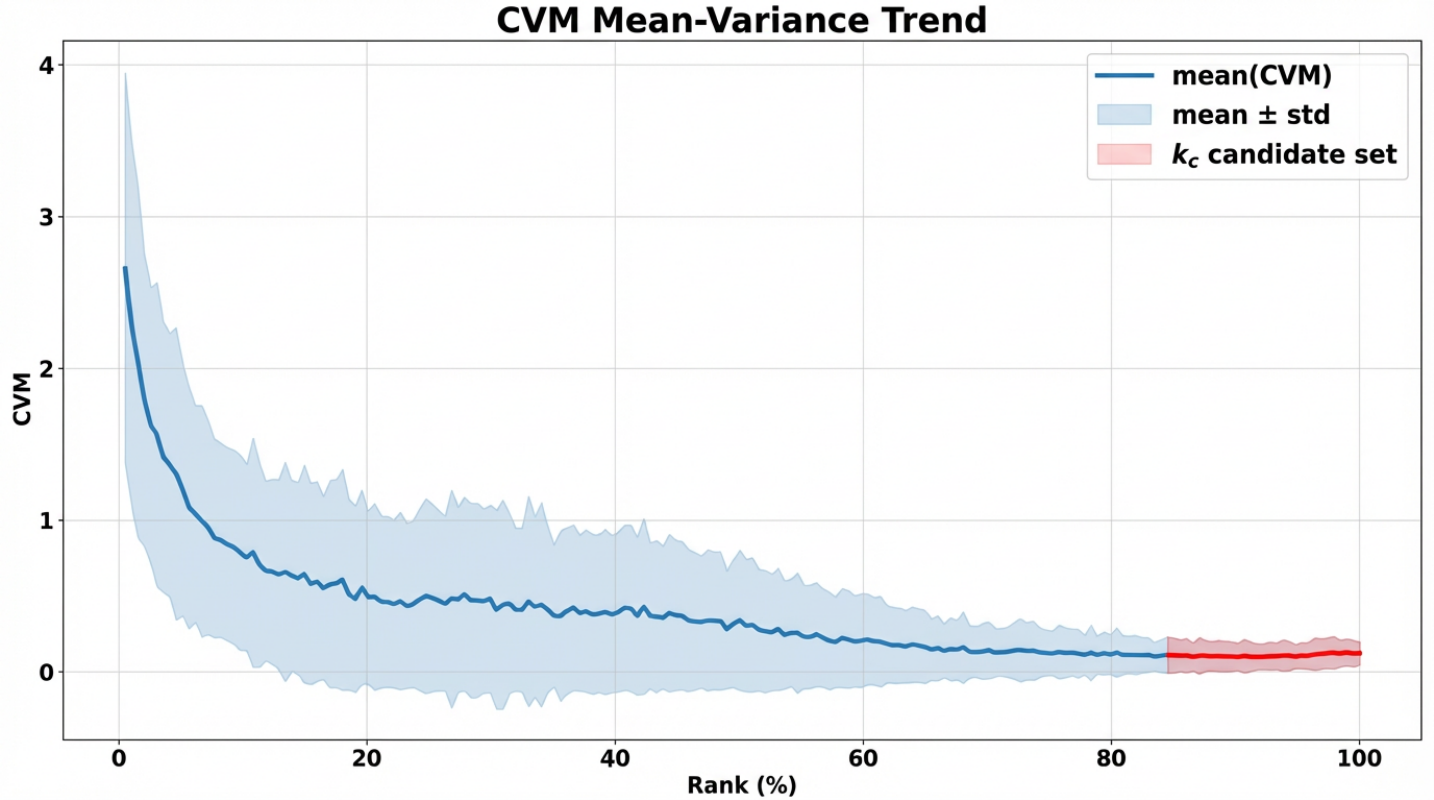}
    \small (b)
  \end{minipage}
  \caption{Empirical validation of the tail stability behavior of the goodness-of-fit statistic $\mathrm{CVM}(k)$.
(a) CVM curves across queries as a function of the retained tail proportion $k$.
For small $k$, the curves exhibit large fluctuations due to contamination from relevant samples.
As $k$ increases, the curves gradually stabilize as the tail becomes dominated by noise.
(b) Mean and standard deviation of $\mathrm{CVM}(k)$ across queries.
The statistic decreases rapidly and then enters a stable regime.
The highlighted region indicates candidate values of the transition point $k_c$,
corresponding to the earliest stable minimum of $\mathrm{CVM}(k)$.}
  \label{fig:cvm}
\end{figure}

\begin{proposition}[Tail Stability Behavior]
Under Assumption 1 and standard EVT regularity conditions,
consider the goodness-of-fit statistic $\mathrm{CVM}(k)$
computed on the tail set $T_k$.

Then the following qualitative behavior holds:

1. When $k < k_c$, the tail samples contain a non-vanishing
fraction of observations from the relevance distribution $p_r$,
which induces deviations from the GPD family and
leads to relatively large goodness-of-fit statistics.

2. When $k \ge k_c$, the tail samples are dominated by the noise
distribution $p_t$. In this regime, EVT theory implies that the
conditional tail distribution is well approximated by a GPD,
and the fitted goodness-of-fit statistics tend to stabilize
under further truncation.

Consequently, the onset of the noise-dominated regime
is typically associated with the earliest stable minimum
of $\mathrm{CVM}(k)$.
\end{proposition}

A weak sufficient condition for the stabilization of $\mathrm{CVM}(k)$ after the
transition point $k_c$ is given in Appendix~\ref{proposition}, which formalizes the intuition that
tail diagnostics become stable once the tail is dominated by noise samples.
Empirically, we examine the behavior of the goodness-of-fit statistic
$\mathrm{CVM}(k)$ under progressive tail truncation. As shown in
Fig.~\ref{fig:cvm}(a), CVM values exhibit large fluctuations for small $k$ but
gradually stabilize as noise dominance increases. Figure~\ref{fig:cvm}(b) further
shows that the mean CVM decreases and enters a stable regime, where the earliest
stable minimum serves as an empirical indicator of the transition point $k_c$.


\subsection{Geometric Localization for Search Reduction}

Evaluating all possible $k$ is computationally expensive
and statistically unstable.
We therefore localize the search region
using geometric knee detection.

Define the deviation from the diagonal:
\begin{equation}
d_i = \frac{y_i - x_i}{\sqrt{2}}.
\end{equation}
The knee index is:
\begin{equation}
k_{\text{knee}} = \arg\max_i d_i.
\end{equation}

Under monotone likelihood ratio mixtures, the regime transition induces maximal curvature in the normalized ranked curve, and the resulting knee index is used solely as a coarse localization window—not as a consistent estimator of the true change-point $k_c$—which is a common and practical strategy in change-point analysis \cite{truong2020selective}.

We restrict the refinement search to a window $\Delta = \left\lceil \sqrt{N \log N} \right\rceil$,
which grows sublinearly with $N$
while empirically covering the transition point $k_c$ with high probability in practice~\cite{yao1988estimating,bai1997estimation}.


\subsection{Algorithm}
We set the minimum tail size to $n_{\min}=5$ to avoid degenerate or ill-posed GPD estimation in extremely small samples. The sensitivity analysis of EVT-related hyperparameters is presented in the Appendix~\ref{ablation}.
\begin{algorithm}[H]
\caption{Tail-Aware Adaptive-$k$}
\begin{algorithmic}[1]
\Require Sorted similarity scores $S$
\State Normalize ranked curve $(x_i,y_i)$
\State Compute knee index $k_{\text{knee}}$
\State Set search window $\Delta = \lceil \sqrt{N \log N} \rceil$
\For{$k \in [k_{\text{knee}}-\Delta,\;k_{\text{knee}}+\Delta]$}
    \If{$|T_k| \ge n_{\min}$}
        \State Fit GPD to tail $T_k$
        \State Compute $\mathrm{CVM}(k)$
    \Else
        \State Set $\mathrm{CVM}(k)=+\infty$
    \EndIf
\EndFor
\State \textbf{return} $k^* = \arg\min_k \mathrm{CVM}(k)$
\end{algorithmic}
\end{algorithm}


\subsection{Computational Complexity}

Knee detection requires $O(N)$ time.
The refinement stage evaluates
$O(\sqrt{N \log N})$ candidate indices.
The overall complexity is $
O\!\left(
N + \sqrt{N \log N} \cdot M
\right)$,
where $M$ denotes the cost of fitting a GPD.


\section{Experiments}
\subsection{Experimental Setup}

\subsubsection{Datasets and Data Splitting.}
We evaluate our method on three widely adopted benchmarks: the single-hop WebQ \cite{berant2013semantic}, and the multi-hop 2Wiki \cite{ho2020constructing} and MuSi \cite{trivedi2022musique}. To rigorously assess our adaptive truncation mechanism, we follow the data processing protocols established by EDC-2-RAG \cite{li2025efficient}. Specifically, using a standard Wikipedia snapshot partitioned into fixed-length segments as the retrieval corpus, we retrieve a comprehensive pool of candidate documents for each test query. Ground truth relevance labels are then automatically assigned to these candidates via exact string matching against the gold answers.\footnote{\url{https://github.com/Tsinghua-dhy/EDC-2-RAG}}

\subsubsection{Evaluation Metrics.}
We evaluate retrieval and end-to-end performance using four metrics.
Precision~\cite{buckley2004retrieval} and Recall~\cite{baeza1999modern}
measure the relevance of retrieved documents and the coverage of gold-standard evidence, respectively.
F1-score~\cite{powers2020evaluation} summarizes their trade-off.
Answer Accuracy~\cite{es2024ragas} evaluates end-to-end effectiveness,
defined as the fraction of queries for which the LLM produces
a ground-truth-matching answer. Additional details are provided in Appendix~\ref{app:llm-prompts}.

\subsubsection{Implementation Details.} We use \texttt{bailian-text-embedding-v4}\footnote{\url{https://www.aliyun.com/product/bailian}} with compressed dimension $d=64$ as the primary retriever to evaluate truncation sensitivity. Robustness is further assessed across different embedding dimensions and multiple retrievers, including \texttt{bailian} \cite{zhang2025qwen3}, \texttt{BGE} \cite{xiao2024c}, and \texttt{Contriever} \cite{izacard2021unsupervised}. All truncation methods are training-free and evaluated on a single NVIDIA A800 GPU. For the model-based baseline \texttt{Self-Route} \cite{li2024retrieval}, we follow the original setup and use \texttt{GPT-4o} \cite{hurst2024gpt} as the decision agent for dynamic top-$k$ selection, with the temperature set to 0. Due to space limitations, we uniformly state that all experiments were conducted with three independent runs. The results show good stability, with performance variations across runs consistently within 1\%.

\subsection{Experimental Results}

In this section, we present the empirical results to validate the effectiveness, efficiency, and robustness of our proposed method. We specifically aim to answer the following research questions:
\begin{itemize}
    \item \textbf{Q1 (Effectiveness):} Can our proposed \textit{TAA-$k$} strategy outperform existing heuristic and model-based methods in overall retrieval quality?
    \item \textbf{Q2 (Efficiency \& End-to-End Optimality):}Does the geometric prior in TAA-$k$ resolve the computational bottleneck of statistical truncation while maintaining near-optimal boundaries and superior downstream LLM performance?
    \item \textbf{Q3 (Robustness):} Does the proposed method maintain its superiority across different embedding models and varying compression dimensions?
\end{itemize}

\subsubsection{Q1:}\textbf{Can our proposed \textit{TAA-$k$} strategy outperform existing heuristic and model-based methods in overall retrieval quality?}

To answer this, we present a comprehensive comparison of dynamic truncation strategies across three datasets in Table~\ref{tab:main_results}. Specifically, we evaluate our proposed \textit{TAA-$k$} method against existing heuristic and model-based dynamic baselines (\textit{Adaptive-$k$}, \textit{Self-Route}, and \textit{surprise}), as well as the optimal F1-score(\textit{Oracle}).

\begin{table}[ht]
\centering
\caption{
Retrieval performance (Precision, Recall, F1-score) of various dynamic truncation methods across three datasets. Experiments are conducted using the \texttt{Bailian-text-embedding-v4} model at 64 dimensions. We provide a detailed interpretation of the precision behavior of TAA-$k$ in Appendix~\ref{app:precision}.
}
\footnotesize
\begin{tabular}{lccccccccc}
\toprule
\multirow{2}{*}{Method} & \multicolumn{3}{c}{WebQ} & \multicolumn{3}{c}{2Wiki} & \multicolumn{3}{c}{MuSi} \\
\cmidrule(lr){2-4}\cmidrule(lr){5-7}\cmidrule(lr){8-10}
 & Precision & Recall & F1-score & Precision & Recall & F1-score & Precision & Recall & F1-score \\
\midrule
Adaptive-$k$          
  & \underline{54.73} & \underline{76.32} & 51.65 
  & 44.15 & 64.45 & 43.56 
  & 48.35 & 71.25 & 48.12 \\
Self-Route            
  & \textbf{58.39} & 22.08 & 22.96 
  & 49.31 & \underline{82.34} & \underline{56.31} 
  & 50.10 & \underline{85.04} & 58.01 \\
surprise         
  & 52.89 & 65.39 & \underline{57.96} 
  & \textbf{50.75} & 63.32 & 56.01 
  & \textbf{52.88} & 66.86 & \underline{58.74} \\
\textbf{TAA-$k$}         
  & 50.62 & \textbf{94.37} & \textbf{65.86} 
  & \underline{50.60} & \textbf{94.25} & \textbf{65.81} 
  & \underline{50.97} & \textbf{95.10} & \textbf{66.34} \\
\midrule
Oracle                
  & 54.13 & 95.97 & 68.75 
  & 52.20 & 97.34 & 67.79
  & 53.64 & 95.66 & 68.38 \\
\bottomrule
\end{tabular}
\label{tab:main_results}
\end{table}

\noindent\textbf{Overall.} As demonstrated in Table~\ref{tab:main_results}, our proposed \textit{TAA-$k$} strategy generally outperforms existing dynamic methods, achieving performance that is remarkably close to the \textit{Oracle}. Across both single-hop and multi-hop datasets, \textit{TAA-$k$} consistently secures the highest F1-scores. Although highly conservative baselines like \textit{Self-Route} occasionally yield higher Precision by aggressively discarding documents, they do so at a severe expense to Recall. By effectively fusing geometric priors with statistical tail modeling, \textit{TAA-$k$} strikes an optimal balance: it drastically improves Recall while preserving competitive Precision, proving to be the most robust solution for maximizing overall retrieval quality.

\noindent\textbf{Analysis.} The superiority of \textit{TAA-$k$} stems from its explicit modeling of the noise-dominated tail relevance distribution. Gap-based heuristics like \textit{Adaptive-$k$} are overly aggressive in multi-hop tasks, failing to capture secondary evidence \cite{yang2018hotpotqa} and dropping Recall to 64.45\% on 2Wiki. Conversely, methods like \textit{Self-Route} and \textit{surprise} lean towards higher Precision but experience a significant drop in Recall. Rather than aggressive filtering, \textit{TAA-$k$} prioritizes overall context retention to successfully recover subtle signals. By using a geometric prior to guide rigorous statistical tail-peeling, it achieves near-perfect Recall and optimal F1-scores that closely approach the \textit{Oracle} upper bound. This demonstrates that our localized search maintains theoretical rigor without compromising retrieval quality, setting the stage for the massive efficiency gains analyzed in Q2.

\subsubsection{Q2:}\textbf{Does the geometric prior in TAA-$k$ resolve the computational bottleneck of statistical truncation while maintaining near-optimal boundaries and superior downstream LLM performance?}

To answer this, we first evaluate the latency reduction and theoretical optimality of TAA-$k$ through a baseline comparisons and ablation study in Table~\ref{tab:optimality} and Figure~\ref{fig:efficiency}. Then we validate its practical effectiveness by reporting the end-to-end LLM answer accuracy across multiple datasets in Table~\ref{tab:llm_accuracy}.

{
\setlength{\tabcolsep}{6pt}
\begin{table}[h]
\caption{
Optimality analysis. Diff-$k$ denotes the mean absolute error between the predicted $k$ and the Oracle $k$, and $\Delta$F1 denotes the F1-score gap between each method and the Oracle. The Oracle and the Diff-$k$ metric are formally defined in
Appendix~\ref{Orale}.
}
\centering
\footnotesize
\begin{tabular}{lcccccc}
\toprule
\multirow{2}{*}{Method} & \multicolumn{2}{c}{WebQ} & \multicolumn{2}{c}{2Wiki} & \multicolumn{2}{c}{MuSi} \\
\cmidrule(lr){2-3}\cmidrule(lr){4-5}\cmidrule(lr){6-7}
& Diff-$k$ & $\Delta$F1 & Diff-$k$ & $\Delta$F1 & Diff-$k$ & $\Delta$F1 \\
\midrule
Adaptive-$k$   
& \underline{56.55} & 17.10 
& 72.70 & 24.23 
& 63.50 & 20.26 \\
Self-Route     
& 143.74 & 45.61 
& \underline{43.42} & 11.90 
& \underline{43.96} & 10.78 \\
surprise       
& 57.72 & \underline{10.79} 
& 63.23 & \underline{11.78} 
& 55.38 & \underline{9.63} \\
\textbf{TAA-$k$} 
& \textbf{19.30} & \textbf{2.89}  
& \textbf{13.10} & \textbf{1.98}  
& \textbf{17.99} & \textbf{2.04} \\
\bottomrule
\end{tabular}
\label{tab:optimality}
\end{table}
}
\noindent\textbf{Overall.} As demonstrated in Table~\ref{tab:optimality}, TAA-$k$ establishes near-optimal truncation boundaries, achieving the lowest Diff-$k$ and $\Delta$F1 deviations from the Oracle compared to fast heuristic and model-based baselines. The ablation study in Figure~\ref{fig:efficiency} illustrates that incorporating the geometric prior slashes inference latency by an order of magnitude compared to exhaustive statistical methods, fully resolving their computational bottlenecks without sacrificing retrieval precision. Table~\ref{tab:llm_accuracy} confirms the end-to-end superiority of this approach, where the high-quality contexts preserved by TAA-$k$ yield the highest average downstream LLM answer accuracy across all evaluated datasets.

\begin{table}[h]
\centering
\begin{minipage}{0.46\linewidth}
    \centering
    \includegraphics[width=\linewidth]{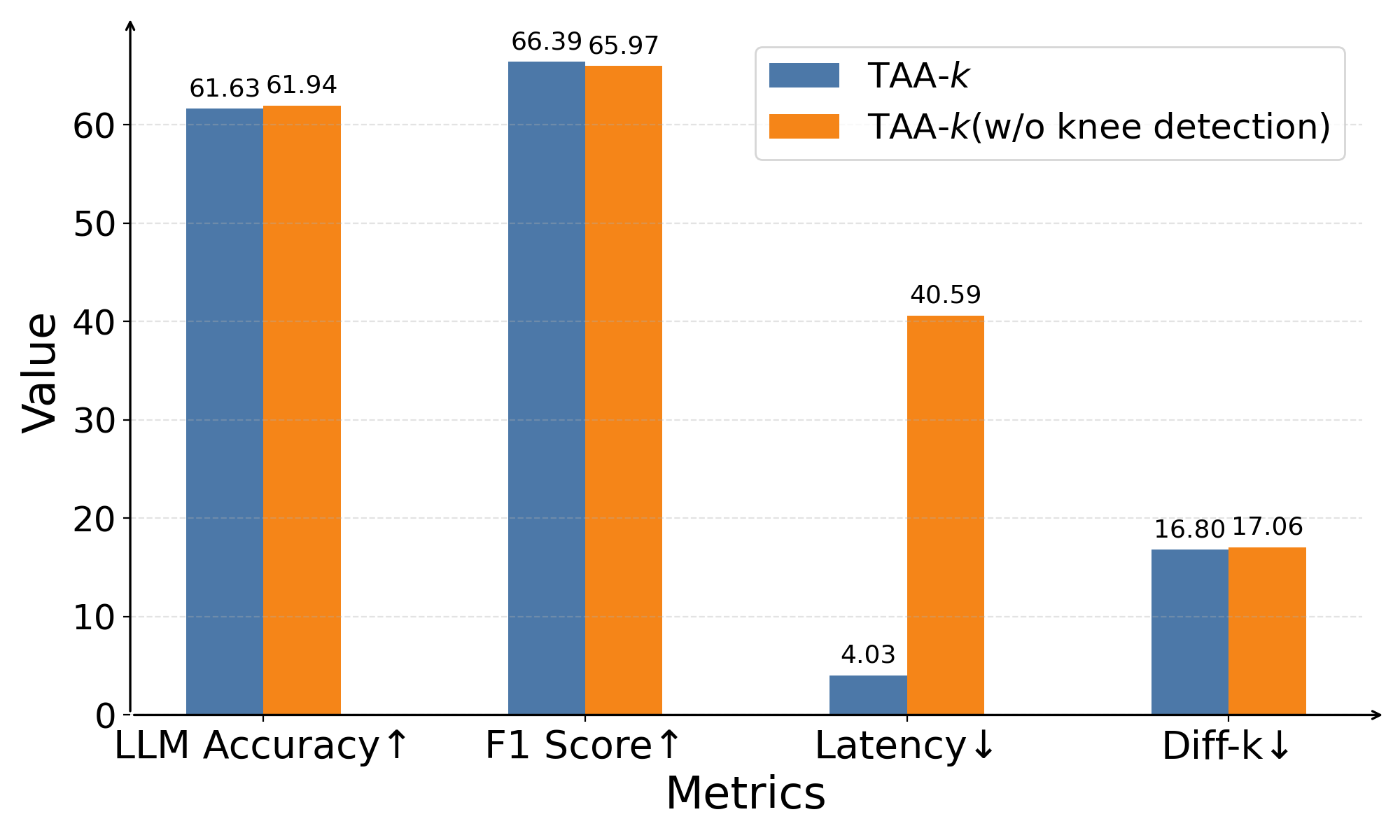}
    \captionof{figure}{Comparison of \textit{TAA-$k$} and method without knee detection across varying evaluation metrics.} 
    \label{fig:efficiency}
\end{minipage}%
\hfill
\begin{minipage}{0.52\linewidth}
    \centering
    \caption{LLM answer accuracy based on documents selected by different dynamic truncation strategies across three datasets.}
    \footnotesize
    \begin{tabular}{lcccc}
    \toprule
    \multirow{2}{*}{Method} & \multicolumn{4}{c}{Answer Accuracy (\%)} \\
    \cmidrule(lr){2-5}
    & WebQ & 2Wiki & MuSi & Avg \\
    \midrule
    Adaptive-$k$      & \underline{78.18} & 47.45        & \underline{47.62} & 57.75 \\
    Self-Route        & 76.91            & 55.56        & \textbf{48.57}    & \underline{60.35} \\
    surprise          & \textbf{79.66}   & \underline{56.71} & 41.90        & 59.42 \\
    \textbf{TAA-$k$}         & \textbf{79.66}   & \textbf{58.56}    & 46.67        & \textbf{61.63} \\
    \bottomrule
    \end{tabular}
    \label{tab:llm_accuracy}
\end{minipage}
\end{table}

\noindent\textbf{Analysis.} In terms of retrieval precision, TAA-$k$ establishes near-optimal boundaries by consistently achieving the lowest $\Delta$F1 gap to the Oracle and the smallest Diff-$k$ error across all datasets. In stark contrast, heuristics like \textit{Adaptive-$k$} and model-based methods like \textit{Self-Route} suffer massive deviations, such as a staggering Diff-$k$ of 143.74 for \textit{Self-Route} on WebQ. Furthermore, Figure~\ref{fig:efficiency} demonstrates that introducing the geometric prior improves retrieval quality over the exhaustive method by actively filtering extreme outliers to increase the F1 Score and reduce Diff-$k$ errors.

The ablation study highlights the geometric prior's critical role in resolving the computational bottleneck of rigorous statistical truncation. The global exhaustive approach suffers from a prohibitive $O(N^2)$ complexity without knee detection. By leveraging the geometric prior to tightly bound the search space, TAA-$k$ drastically reduces this dominating complexity to $O(\sqrt{N \log N})$. As Figure~\ref{fig:efficiency} validates, this localized coarse-to-fine design \cite{liu2017cascade} slashes inference latency from 40.59 ms to 4.03 ms, achieving a 10$\times$ speedup while maintaining strictly comparable downstream LLM answer accuracy to the exhaustive method. Moreover, while model-based approaches like \textit{Self-Route} rely on high-latency, cost-prohibitive LLM API calls \cite{chen2023frugalgpt} to determine the truncation position, TAA-$k$ operates training-free and resolves the truncation in mere milliseconds without external dependencies.

In terms of downstream reasoning, TAA-$k$ offers the most effective context reduction strategy. Aggressive heuristics achieve token reduction by indiscriminately discarding documents, severely penalizing generation quality since language models are highly sensitive to missing or fragmented context \cite{shi2023large}. As Table~\ref{tab:llm_accuracy} demonstrates, TAA-$k$ preserves evidence chain integrity, securing the highest average LLM answer accuracy among dynamic methods. Notably, it outperforms strong baselines like \textit{Self-Route} on complex multi-hop datasets like 2Wiki at 58.56\% and ties for the top score on WebQ, proving that TAA-$k$'s precise boundaries effectively enhance LLM reasoning capabilities.

\subsubsection{Q3:}
\textbf{Does the proposed method maintain its superiority across different embedding models and varying compression dimensions?}

To answer this, we conduct extensive experiments to verify the universality of our approach across diverse dense retriever architectures (Contriever, BGE, and Qwen3) under varying embedding dimensions ranging from $d=64$ to $1024$. The comprehensive F1-score comparison is detailed in Table~\ref{tab:robustness}.

\begin{table*}[h]
\caption{
\textbf{Robustness Analysis.} F1-score comparison of dynamic Top-$k$ strategies across datasets, retrievers, and embedding dimensions. Rows enumerate datasets (WebQ, 2Wiki, MuSiQue) and dynamic methods. Columns correspond to three dense retrievers (Contriever at 768d, BGE, Qwen3) evaluated at varying dimensions.
}
\centering
\footnotesize
\setlength{\tabcolsep}{3pt}
\begin{tabular}{cccccccccc}
\toprule
\multicolumn{2}{c}{\multirow{2}{*}{Method}} &
\multicolumn{1}{c}{Contriever} &
\multicolumn{3}{c}{BGE} &
\multicolumn{4}{c}{Qwen3} \\
\cmidrule(lr){3-3}\cmidrule(lr){4-6}\cmidrule(lr){7-10}
\multicolumn{2}{c}{} &
768 &
384 & 768 & 1024 &
64 & 256 & 768 & 1024 \\
\midrule
\multirow{4}{*}{WebQ}
& Adaptive-$k$    
& 34.69 & 34.33 & 33.41 & 33.66 & 51.65 & 44.99 & 37.91 & 37.83 \\
& Self-Route      
& 23.46 & 24.96 & 24.00 & 24.56 & 22.96 & 23.62 & 24.28 & 24.37 \\
& Surprise        
& \underline{54.86} & \underline{58.46} & \underline{58.70} & \underline{59.69} 
& \underline{57.96} & \underline{58.13} & \underline{57.70} & \underline{57.83} \\
& \textbf{TAA-$k$}
& \textbf{65.53} & \textbf{66.21} & \textbf{66.40} & \textbf{66.63} 
& \textbf{65.86} & \textbf{66.17} & \textbf{66.20} & \textbf{66.32} \\
\midrule
\multirow{4}{*}{2Wiki}
& Adaptive-$k$    
& 18.55 & 21.71 & 23.18 & 22.52 & 43.56 & 30.03 & 22.02 & 22.78 \\
& Self-Route      
& \underline{56.03} & 55.87 & 55.26 & 55.23 
& \underline{56.31} & 55.50 & 55.66 & 55.42 \\
& Surprise        
& 53.30 & \underline{56.49} & \underline{56.25} & \underline{57.03} 
& 56.01 & \underline{55.91} & \underline{55.81} & \underline{55.59} \\
& \textbf{TAA-$k$}
& \textbf{65.58} & \textbf{66.28} & \textbf{66.18} & \textbf{66.58} 
& \textbf{65.81} & \textbf{66.06} & \textbf{66.01} & \textbf{66.02} \\
\midrule
\multirow{4}{*}{MuSiQue}
& Adaptive-$k$    
& 28.62 & 35.05 & 28.26 & 33.89 & 48.12 & 41.19 & 29.96 & 26.99 \\
& Self-Route      
& \underline{55.55} & 54.13 & 55.93 & 56.46 
& 58.01 & 53.81 & 54.96 & 54.90 \\
& Surprise        
& 54.92 & \underline{57.67} & \underline{58.02} & \underline{58.69} 
& \underline{58.74} & \underline{57.49} & \underline{58.13} & \underline{58.15} \\
& \textbf{TAA-$k$}
& \textbf{65.62} & \textbf{66.42} & \textbf{66.29} & \textbf{66.47} 
& \textbf{66.34} & \textbf{66.18} & \textbf{66.37} & \textbf{66.40} \\
\bottomrule
\end{tabular}
\label{tab:robustness}
\end{table*}

\noindent\textbf{Overall.} As evidenced by Table~\ref{tab:robustness}, our \textit{TAA-$k$} algorithm exhibits remarkable robustness, consistently outperforming baseline methods across all tested environments. Whether applied to unsupervised dense models like Contriever or instruction-tuned models such as BGE and Qwen3, and regardless of whether the embeddings are full-dimensional at $d=1024$ or heavily compressed to $d=64$, \textit{TAA-$k$} maintains the highest F1-scores. This demonstrates that it serves as a universal, plug-and-play module for dynamic retrieval pipelines.

\noindent\textbf{Analysis.} The stability of our framework highlights two critical properties. Regarding dimension insensitivity, heuristic methods like \textit{Adaptive-$k$} suffer severe performance fluctuations across different compression levels. This likely occurs because absolute score gaps become distorted in heavily compressed spaces \cite{kusupati2022matryoshka}, exacerbating representation anisotropy \cite{ethayarajh2019contextual} and causing heuristics to misfire. In contrast, \textit{TAA-$k$} remains remarkably stable regardless of embedding dimensions, suggesting our statistical modeling captures a fundamental geometric property of relevance ranking that persists even under severe semantic compression.

Regarding model agnosticism, \textit{TAA-$k$} adapts seamlessly to distinct score distributions. Different architectures inherently produce varying similarity scales and space uniformities \cite{wang2020understanding}; for example, unsupervised contrastive models (e.g., Contriever) yield different score densities than instruction-tuned models (e.g., BGE, Qwen3). Despite this, \textit{TAA-$k$} consistently identifies the optimal cut-off across all evaluated retrievers. This indicates our strategy relies on the relative curvature and statistical tail properties of the ranking list rather than absolute score values, eliminating the need for model-specific hyperparameter tuning and ensuring robustness across diverse retrieval backbones.

\section{Discussion and Conclusion}

In this paper, we introduced TAA-$k$, a novel coarse-to-fine adaptive context truncation framework for long-context question answering. Unlike existing heuristic or computationally expensive model-based methods, TAA-$k$ fundamentally resolves the efficiency-precision tradeoff by fusing a geometric prior with rigorous extreme value theory. By initially leveraging knee detection to tightly bound the search space and subsequently modeling the tail of the similarity score distribution with a generalized Pareto distribution, our method accurately identifies the transition to the noise-dominated regime. Extensive empirical evaluations demonstrate that TAA-$k$ achieves a remarkable 10$\times$ reduction in inference latency (reducing complexity to $O(\sqrt{N \log N} \cdot M)$) while maintaining near-optimal boundaries that closely approach the Oracle. Furthermore, it serves as a robust plug-and-play module that exhibits extreme resilience to heavily compressed embedding spaces and diverse retriever architectures, ultimately delivering the highest downstream LLM reasoning accuracy.

Although heavy-tail behavior is commonly observed in retrieval distributions \cite{clauset2009power,resnick2007heavy}, severe score overlap or exceptionally small candidate pools can blur the relevance-noise separation and induce finite-sample effects in GPD estimation \cite{coles2001introduction}. We effectively mitigate these statistical constraints within our current framework by enforcing a minimum tail size threshold and tracking the earliest stable minimum of the CVM curve. To further enhance robustness, future work will explore integrating this adaptive threshold estimation directly into retrieval or reranking models, enabling joint modeling of retrieval uncertainty and context utilization in end-to-end retrieval-augmented generation systems.

\begin{credits}
\subsubsection{\ackname} The authors would like to thank the anonymous reviewers for their helpful comments.

\subsubsection{\discintname}
The authors have no competing interests to declare that are relevant to the content of this article.
\end{credits}
%
%
%
%
\bibliographystyle{splncs04}
\bibliography{references}

@article{lewis2020retrieval,
  title={Retrieval-augmented generation for knowledge-intensive nlp tasks},
  author={Lewis, Patrick and Perez, Ethan and Piktus, Aleksandra and Petroni, Fabio and Karpukhin, Vladimir and Goyal, Naman and K{\"u}ttler, Heinrich and Lewis, Mike and Yih, Wen-tau and Rockt{\"a}schel, Tim and others},
  journal={Advances in neural information processing systems},
  volume={33},
  pages={9459--9474},
  year={2020}
}

@article{gao2023retrieval,
  title={Retrieval-augmented generation for large language models: A survey},
  author={Gao, Yunfan and Xiong, Yun and Gao, Xinyu and Jia, Kangxiang and Pan, Jinliu and Bi, Yuxi and Dai, Yixin and Sun, Jiawei and Wang, Haofen and Wang, Haofen},
  journal={arXiv preprint arXiv:2312.10997},
  volume={2},
  number={1},
  year={2023}
}

@inproceedings{taguchi-etal-2025-efficient,
    title = "Efficient Context Selection for Long-Context {QA}: No Tuning, No Iteration, Just Adaptive{-}$k$",
    author = "Taguchi, Chihiro  and
      Maekawa, Seiji  and
      Bhutani, Nikita",
    booktitle = "Proceedings of the 2025 Conference on Empirical Methods in Natural Language Processing",
    pages = {20105--20130},
    year = {2025},
    publisher = "Association for Computational Linguistics",
}

@inproceedings{karpukhin2020dense,
  title={Dense Passage Retrieval for Open-Domain Question Answering},
  author={Karpukhin, Vladimir and Oguz, Barlas and Min, Sewon and Lewis, Patrick and Wu, Ledell and Edunov, Sergey and Chen, Danqi and Yih, Wen-tau},
  booktitle={Proceedings of the 2020 Conference on Empirical Methods in Natural Language Processing (EMNLP)},
  pages={6769--6781},
  year={2020}
}

@article{robertson2009probabilistic,
  title={The Probabilistic Relevance Framework: BM25 and Beyond},
  author={Robertson, Stephen and Zaragoza, Hugo},
  journal={Information Retrieval},
  volume={3},
  number={4},
  pages={333--389},
  year={2009}
}

@inproceedings{sawarkar2024blended,
  title={Blended rag: Improving rag (retriever-augmented generation) accuracy with semantic search and hybrid query-based retrievers},
  author={Sawarkar, Kunal and Mangal, Abhilasha and Solanki, Shivam Raj},
  booktitle={2024 IEEE 7th international conference on multimedia information processing and retrieval (MIPR)},
  pages={155--161},
  year={2024},
  organization={IEEE}
}

@inproceedings{berant2013semantic,
  title={Semantic parsing on freebase from question-answer pairs},
  author={Berant, Jonathan and Chou, Andrew and Frostig, Roy and Liang, Percy},
  booktitle={Proceedings of the 2013 conference on empirical methods in natural language processing},
  pages={1533--1544},
  year={2013}
}

@inproceedings{ho2020constructing,
  title={Constructing a multi-hop qa dataset for comprehensive evaluation of reasoning steps},
  author={Ho, Xanh and Nguyen, Anh-Khoa Duong and Sugawara, Saku and Aizawa, Akiko},
  booktitle={Proceedings of the 28th International Conference on Computational Linguistics},
  pages={6609--6625},
  year={2020}
}

@article{trivedi2022musique,
  title={MuSiQue: Multihop Questions via Single-hop Question Composition},
  author={Trivedi, Harsh and Balasubramanian, Niranjan and Khot, Tushar and Sabharwal, Ashish},
  journal={Transactions of the Association for Computational Linguistics},
  volume={10},
  pages={539--554},
  year={2022},
  publisher={MIT Press One Broadway, 12th Floor, Cambridge, Massachusetts 02142, USA~…}
}

@article{li2025efficient,
  title={Efficient dynamic clustering-based document compression for retrieval-augmented-generation},
  author={Li, Weitao and Liu, Kaiming and Zhang, Xiangyu and Lei, Xuanyu and Ma, Weizhi and Liu, Yang},
  journal={arXiv preprint arXiv:2504.03165},
  year={2025}
}

@inproceedings{xiao2024c,
  title={C-pack: Packed resources for general chinese embeddings},
  author={Xiao, Shitao and Liu, Zheng and Zhang, Peitian and Muennighoff, Niklas and Lian, Defu and Nie, Jian-Yun},
  booktitle={Proceedings of the 47th international ACM SIGIR conference on research and development in information retrieval},
  pages={641--649},
  year={2024}
}

@inproceedings{li2024retrieval,
  title={Retrieval augmented generation or long-context llms? a comprehensive study and hybrid approach},
  author={Li, Zhuowan and Li, Cheng and Zhang, Mingyang and Mei, Qiaozhu and Bendersky, Michael},
  booktitle={Proceedings of the 2024 Conference on Empirical Methods in Natural Language Processing: Industry Track},
  pages={881--893},
  year={2024}
}

@article{liu2024lost,
  title={Lost in the middle: How language models use long contexts},
  author={Liu, Nelson F and Lin, Kevin and Hewitt, John and Paranjape, Ashwin and Bevilacqua, Michele and Petroni, Fabio and Liang, Percy},
  journal={Transactions of the association for computational linguistics},
  volume={12},
  pages={157--173},
  year={2024}
}

@inproceedings{bahri2023surprise,
  title={Surprise: Result list truncation via extreme value theory},
  author={Bahri, Dara and Zheng, Che and Tay, Yi and Metzler, Donald and Tomkins, Andrew},
  booktitle={Proceedings of the 46th International ACM SIGIR Conference on Research and Development in Information Retrieval},
  pages={2404--2408},
  year={2023}
}

@article{zhang2025qwen3,
  title={Qwen3 embedding: Advancing text embedding and reranking through foundation models},
  author={Zhang, Yanzhao and Li, Mingxin and Long, Dingkun and Zhang, Xin and Lin, Huan and Yang, Baosong and Xie, Pengjun and Yang, An and Liu, Dayiheng and Lin, Junyang and others},
  journal={arXiv preprint arXiv:2506.05176},
  year={2025}
}

@article{izacard2021unsupervised,
  title={Unsupervised dense information retrieval with contrastive learning},
  author={Izacard, Gautier and Caron, Mathilde and Hosseini, Lucas and Riedel, Sebastian and Bojanowski, Piotr and Joulin, Armand and Grave, Edouard},
  journal={arXiv preprint arXiv:2112.09118},
  year={2021}
}

@article{hurst2024gpt,
  title={Gpt-4o system card},
  author={Hurst, Aaron and Lerer, Adam and Goucher, Adam P and Perelman, Adam and Ramesh, Aditya and Clark, Aidan and Ostrow, AJ and Welihinda, Akila and Hayes, Alan and Radford, Alec and others},
  journal={arXiv preprint arXiv:2410.21276},
  year={2024}
}

@inproceedings{jiang2023active,
  title={Active retrieval augmented generation},
  author={Jiang, Zhengbao and Xu, Frank F and Gao, Luyu and Sun, Zhiqing and Liu, Qian and Dwivedi-Yu, Jane and Yang, Yiming and Callan, Jamie and Neubig, Graham},
  booktitle={Proceedings of the 2023 conference on empirical methods in natural language processing},
  pages={7969--7992},
  year={2023}
}

@article{pickands1975statistical,
  title={Statistical inference using extreme order statistics},
  author={Pickands III, James},
  journal={the Annals of Statistics},
  pages={119--131},
  year={1975},
  publisher={JSTOR}
}

@book{coles2001introduction,
  title={An introduction to statistical modeling of extreme values},
  author={Coles, Stuart and Bawa, Joanna and Trenner, Lesley and Dorazio, Pat},
  volume={208},
  year={2001},
  publisher={Springer}
}

@book{embrechts2013modelling,
  title={Modelling extremal events: for insurance and finance},
  author={Embrechts, Paul and Kl{\"u}ppelberg, Claudia and Mikosch, Thomas},
  volume={33},
  year={2013},
  publisher={Springer Science \& Business Media}
}

@book{cramer1928composition,
  title={On the composition of elementary errors: Statistical applications},
  author={Cram{\'e}r, Harald},
  year={1928},
  publisher={Almqvist and Wiksell}
}

@article{anderson1952asymptotic,
  title={Asymptotic theory of certain" goodness of fit" criteria based on stochastic processes},
  author={Anderson, Theodore W and Darling, Donald A},
  journal={The annals of mathematical statistics},
  pages={193--212},
  year={1952},
  publisher={JSTOR}
}

@article{yao1988estimating,
  title={Estimating the number of change-points via Schwarz'criterion},
  author={Yao, Yi-Ching},
  journal={Statistics \& Probability Letters},
  volume={6},
  number={3},
  pages={181--189},
  year={1988},
  publisher={Elsevier}
}

@article{bai1997estimation,
  title={Estimation of a change point in multiple regression models},
  author={Bai, Jushan},
  journal={Review of Economics and Statistics},
  volume={79},
  number={4},
  pages={551--563},
  year={1997},
}

@book{resnick2007heavy,
  title={Heavy-tail phenomena: probabilistic and statistical modeling},
  author={Resnick, Sidney I},
  year={2007},
  publisher={Springer}
}

@article{clauset2009power,
  title={Power-law distributions in empirical data},
  author={Clauset, Aaron and Shalizi, Cosma Rohilla and Newman, Mark EJ},
  journal={SIAM review},
  volume={51},
  number={4},
  pages={661--703},
  year={2009},
  publisher={SIAM}
}

@article{truong2020selective,
  title={Selective review of offline change point detection methods},
  author={Truong, Charles and Oudre, Laurent and Vayatis, Nicolas},
  journal={Signal processing},
  volume={167},
  pages={107299},
  year={2020},
  publisher={Elsevier}
}

@inproceedings{buckley2004retrieval,
  title={Retrieval evaluation with incomplete information},
  author={Buckley, Chris and Voorhees, Ellen M},
  booktitle={Proceedings of the 27th annual international ACM SIGIR conference on Research and development in information retrieval},
  pages={25--32},
  year={2004}
}

@book{baeza1999modern,
  title={Modern information retrieval},
  author={Baeza-Yates, Ricardo and Ribeiro-Neto, Berthier and others},
  volume={463},
  year={1999},
  publisher={ACM press New York}
}

@article{powers2020evaluation,
  title={Evaluation: from precision, recall and F-measure to ROC, informedness, markedness and correlation},
  author={Powers, David MW},
  journal={arXiv preprint arXiv:2010.16061},
  year={2020}
}

@article{kusupati2022matryoshka,
  title={Matryoshka representation learning},
  author={Kusupati, Aditya and Bhatt, Gantavya and Rege, Aniket and Wallingford, Matthew and Sinha, Aditya and Ramanujan, Vivek and Howard-Snyder, William and Chen, Kaifeng and Kakade, Sham and Jain, Prateek and others},
  journal={Advances in Neural Information Processing Systems},
  volume={35},
  pages={30233--30249},
  year={2022}
}

@inproceedings{yang2018hotpotqa,
  title={HotpotQA: A dataset for diverse, explainable multi-hop question answering},
  author={Yang, Zhilin and Qi, Peng and Zhang, Saizheng and Bengio, Yoshua and Cohen, William and Salakhutdinov, Ruslan and Manning, Christopher D},
  booktitle={Proceedings of the 2018 conference on empirical methods in natural language processing},
  pages={2369--2380},
  year={2018}
}

@inproceedings{shi2023large,
  title={Large language models can be easily distracted by irrelevant context},
  author={Shi, Freda and Chen, Xinyun and Misra, Kanishka and Scales, Nathan and Dohan, David and Chi, Ed H and Sch{\"a}rli, Nathanael and Zhou, Denny},
  booktitle={International Conference on Machine Learning},
  pages={31210--31227},
  year={2023},
  organization={PMLR}
}

@inproceedings{liu2017cascade,
  title={Cascade ranking for operational e-commerce search},
  author={Liu, Shichen and Xiao, Fei and Ou, Wenwu and Si, Luo},
  booktitle={Proceedings of the 23rd ACM SIGKDD International Conference on Knowledge Discovery and Data Mining},
  pages={1557--1565},
  year={2017}
}

@inproceedings{ethayarajh2019contextual,
  title={How contextual are contextualized word representations? Comparing the geometry of BERT, ELMo, and GPT-2 embeddings},
  author={Ethayarajh, Kawin},
  booktitle={Proceedings of the 2019 conference on empirical methods in natural language processing and the 9th international joint conference on natural language processing (EMNLP-IJCNLP)},
  pages={55--65},
  year={2019}
}

@inproceedings{wang2020understanding,
  title={Understanding contrastive representation learning through alignment and uniformity on the hypersphere},
  author={Wang, Tongzhou and Isola, Phillip},
  booktitle={International conference on machine learning},
  pages={9929--9939},
  year={2020},
  organization={PMLR}
}

@inproceedings{es2024ragas,
  title={Ragas: Automated evaluation of retrieval augmented generation},
  author={Es, Shahul and James, Jithin and Anke, Luis Espinosa and Schockaert, Steven},
  booktitle={Proceedings of the 18th conference of the european chapter of the association for computational linguistics: system demonstrations},
  pages={150--158},
  year={2024}
}

@article{chen2023frugalgpt,
  title={Frugalgpt: How to use large language models while reducing cost and improving performance},
  author={Chen, Lingjiao and Zaharia, Matei and Zou, James},
  journal={arXiv preprint arXiv:2305.05176},
  year={2023}
}

\clearpage
\appendix
\section{Additional Analysis of Tail Stability Criterion}\label{proposition}

We provide a weak sufficient condition under which the goodness-of-fit statistic
$\mathrm{CVM}(k)$ exhibits stabilization after the transition point $k_c$.

\paragraph{Setting.}
Assume the similarity scores are generated from a mixture
$p(s) = \pi_r p_r(s) + (1-\pi_r)p_t(s)$, where $p_t$ belongs to the Weibull domain
of attraction with finite lower endpoint.

\paragraph{Proposition A.1 (Heuristic Stability).}
If there exists $k_c$ such that for all $k \ge k_c$,
$$\mathbb{P}(s_i \sim p_r \mid i > k) \le \epsilon,
$$for sufficiently small $\epsilon > 0$, then the fitted GPD parameters
$(\hat{\xi}_k, \hat{\sigma}_k)$ vary at most $O(\epsilon)$ under further truncation,
and the CVM statistic satisfies
$$\mathbb{E}|\mathrm{CVM}(k+1) - \mathrm{CVM}(k)| = O(\epsilon).
$$
\paragraph{Discussion.}
This result does not rely on asymptotic convergence, but formalizes the notion
that once relevance contamination becomes negligible, tail diagnostics stabilize.

\section{Oracle Truncation and Diff-$k$ Definition}\label{Orale}

Let $\mathcal{Q}$ denote the set of evaluation queries.
For each query $q \in \mathcal{Q}$, the retriever returns a ranked list of documents
$\{d_{q,1}, d_{q,2}, \dots, d_{q,N}\}$ sorted by decreasing similarity score.

\paragraph{Oracle Truncation Index.}
Given ground-truth relevance labels for query $q$, we define the oracle truncation
index $k^{\star}_{\text{oracle}}(q)$ as the prefix length that maximizes the
retrieval F1-score:
$$k^{\star}_{\text{oracle}}(q)
=
\arg\max_{1 \le k \le N}
\mathrm{F1}\!\left(\{d_{q,1}, \dots, d_{q,k}\}\right).
$$This oracle is query-dependent and is only used for retrospective evaluation,
not available at inference time.

\paragraph{Diff-$k$ Metric.}
For a given adaptive truncation method that outputs a truncation index $k(q)$
for each query $q$, we define Diff-$k$ as the mean absolute deviation from the
oracle truncation:
$$\mathrm{Diff}\text{-}k
=
\mathbb{E}_{q \sim \mathcal{Q}}
\left[
\left| k(q) - k^{\star}_{\text{oracle}}(q) \right|
\right],
$$where the expectation is approximated by the empirical average over all queries
in $\mathcal{Q}$.

\paragraph{Discussion.}
Diff-$k$ measures how closely an adaptive truncation strategy approximates the
per-query optimal truncation length. A lower Diff-$k$ indicates better alignment
with the oracle choice.

\section{Sensitivity Analysis of EVT-related Hyperparameters.}\label{ablation}
\begin{table*}[t]
\centering
\caption{Sensitivity analysis of EVT-related hyperparameters in TAA-k.}
\begin{subtable}[t]{0.48\textwidth}
\centering
\caption{Sensitivity to tail window size $\Delta$}
\begin{tabular}{lccc}
\toprule
Window $\Delta$ & Precision & Recall & F1 \\
\midrule
0 (Knee)           & 50.43 & 88.89 & 64.30 \\
$\lceil \sqrt{N} \rceil$       & 50.78 & 91.07 & 65.15 \\
$\lceil \sqrt{N\log N} \rceil$ & 50.60 & 94.25 & 65.81 \\
\bottomrule
\label{hparametersa}
\end{tabular}
\end{subtable}
\hfill
\begin{subtable}[t]{0.48\textwidth}
\centering
\caption{Sensitivity to minimum tail size $n_{\min}$}
\begin{tabular}{lccc}
\toprule
$n_{\min}$ & Precision & Recall & F1 \\
\midrule
5  & 50.60 & 94.25 & 65.81 \\
10 & 50.84 & 87.42 & 64.24 \\
15 & 50.89 & 86.04 & 63.91 \\
\bottomrule
\label{hparametersb}
\end{tabular}
\end{subtable}
\label{hparameters}
\end{table*}

Table~\ref{hparameters} reports the sensitivity of TAA-k to key EVT-related hyperparameters, including the tail window size $\Delta$ and the minimum tail size $n_{\min}$.

As shown in Table~\ref{hparametersa}, different strategies for determining the tail window size yield similar precision values, while recall consistently improves as the window size increases. In particular, using $\Delta=\lceil \sqrt{N\log N} \rceil$ achieves the highest recall (94.25\%) and the best overall F1 score (65.81), indicating that a larger tail window enables the EVT model to better capture extreme behaviors without sacrificing precision.

Table~\ref{hparametersb} examines the effect of varying the minimum tail size $n_{\min}$. When $n_{\min}=5$, the model attains the highest recall and F1 score. As $n_{\min}$ increases, recall and overall performance gradually decrease. This suggests that enforcing a larger minimum tail size may exclude informative extreme samples, thereby weakening the effectiveness of EVT-based modeling.

Overall, the results demonstrate that TAA-k is relatively robust to reasonable choices of EVT hyperparameters. Based on these observations, we adopt $\Delta=\lceil \sqrt{N\log N} \rceil$ and $n_{\min}=5$ as the default settings in subsequent experiments.

\section{Interpretation of Precision Behavior in Adaptive Truncation}
\label{app:precision}

In Table~\ref{tab:main_results}, TAA-$k$ exhibits slightly lower document-level precision compared to some adaptive truncation baselines. In this section, we clarify that this behavior is an inherent and deliberate consequence of recall-oriented truncation, rather than a deficiency of the proposed criterion.

\subsection{Recall-Oriented Truncation and Borderline Documents}

Unlike gap-based or heuristic truncation methods that aggressively prune documents based on local score differences, TAA-$k$ is explicitly designed to detect the \emph{earliest noise-dominated regime} in the ranked list. As a result, the selected cutoff often lies close to the relevance--noise transition boundary, where documents are weakly relevant or partially informative.

Under strict binary relevance annotation, such borderline documents are labeled as non-relevant, which naturally reduces measured precision. However, retaining these documents is essential for achieving high recall and for preserving supporting evidence that may not be captured by binary relevance labels.

\subsection{Precision vs.\ End-to-End Utility}

Document-level precision does not fully reflect downstream utility in retrieval-augmented generation. In multi-hop or open-domain question answering, documents labeled as non-relevant may still provide partial facts, entity disambiguation cues, or contextual information that facilitates correct reasoning.

This observation is empirically supported by Table~\ref{tab:llm_accuracy}, where TAA-$k$ consistently improves end-to-end QA performance despite lower precision. This indicates that the retained documents, although penalized under precision metrics, contribute positively to downstream task performance.

\subsection{Comparison with Oracle Truncation}

To further contextualize the precision behavior, we compare TAA-$k$ with an oracle truncation strategy that selects $k$ to maximize retrieval F1. We observe that the oracle often attains precision values comparable to or lower than TAA-$k$, suggesting that reduced precision is intrinsic to recall-optimal truncation rather than an artifact of our method.

This result reinforces that TAA-$k$ operates close to the recall-optimal regime, prioritizing coverage of relevant evidence over aggressive pruning.

\subsection{Practical Implications}

Overall, the lower precision of TAA-$k$ should be interpreted as a principled trade-off in favor of recall and downstream task accuracy. For retrieval-augmented generation systems, where missing critical evidence is often more detrimental than including mildly noisy context, this trade-off is empirically and practically justified.

\section{LLM Evaluation Prompts}
\label{app:llm-prompts}

This section documents the exact prompts used for evaluating LLM-based answer
accuracy. All prompts are fixed and applied uniformly across models and datasets
to ensure reproducibility. All evaluations are conducted using GPT-4o with the
temperature parameter strictly set to 0, and each experiment is repeated
three times to verify result consistency.

\subsection{Answer Generation Prompt}
\label{app:answer-prompt}

The following prompt is used to elicit an answer from the language model given a
question and retrieved context:

\begin{tcolorbox}[colback=gray!5!white, colframe=darkgray, boxrule=0.5pt, arc=2mm, left=2mm, right=2mm, top=2mm, bottom=2mm]
\small\ttfamily
Your task is to answer the question provided. To help you answer accurately,
some relevant context documents have been retrieved. After reviewing them,
you'll be asked the same question again. Please respond succinctly.\\[1ex]
Input:\\
- Question:\\
\{question\}\\[1ex]
- Context:\\
\{context\}\\[1ex]
- Question:\\
\{question\}\\[1ex]
Response:\\
- Answer:
\end{tcolorbox}

\subsection{Answer Accuracy Evaluation Prompt}
\label{app:eval-prompt}

The following prompt is used to automatically evaluate the correctness of the
model-generated answer. The evaluator is instructed to rely strictly on the
provided response and expected answers, without using external knowledge or
guessing.

\begin{tcolorbox}[colback=gray!5!white, colframe=darkgray, boxrule=0.5pt, arc=2mm, left=2mm, right=2mm, top=2mm, bottom=2mm]
\small\ttfamily
You are a strict evaluator. DO NOT use external knowledge. DO NOT guess.\\
Judge ONLY based on whether the model response contains a correct answer matching ANY of the expected answers.\\[1ex]
Output format (STRICT): output exactly one token: PASS or FAIL. No other words.\\[1ex]
Question: \{question\}\\
Expected answers: \{expected\}\\
Model response: \{response\}
\end{tcolorbox}

\end{document}